\documentstyle[12pt,epsfig]{article}
\begin{document}
\noindent
\begin{center}
{\Large {\bf Hubble Tension in Power-Law $f(R)$ Gravity and Generalized Brans-Dicke Theory }}\\
\vspace{2cm}
 ${\bf Yousef~Bisabr}$\footnote{e-mail:~y-bisabr@sru.ac.ir.}\\
\vspace{.5cm} {\small{Department of Physics, Shahid Rajaee Teacher
Training University,
Lavizan, Tehran 16788, Iran}}\\
\end{center}
\begin{abstract}
We introduce a theoretical framework to alleviate the Hubble tension. This framework is based on dynamics of a minimally coupled scalar field which either belongs to the Brans-Dicke theory with a self-interacting potential or is the scalar partner of $f(R)$ gravity. These two theories are dynamically equivalent when the Brans-Dicke parameter is zero. We will use this dynamical equivalence to interpret the Hubble tension in the same theoretical framework. For both theories, we write one set of field equations in which the value of a parameter distinguishes between the two theories. We will show that $H_0$ actually evolves with redshift so that its value is consistent with that measured from the local distance ladder and it drops to the value measured from CMB at high redshift. We argue that even though both theories exhibit this behaviour, Brans-Dicke theory with an exponential potential is more successful than power-law $f(R)$ gravity to relieve the Hubble tension.

\end{abstract}

~~~~~~~~~~~~~~~~~~~~~~~~~~~~~~~~~~~~~~~~~~~~~~~~~~~~~~~~~~~~~~~~~~~~~~~~~~~~~~~~~~~~~~~~~~~~~~~~~~~~~
\section{Introduction}
There are a number of unexplained features in the standard cosmology among which the accelerating expansion of the Universe has been recently drawn much amount of attention. There is not still a well-understood explanation on the nature of this phenomenon. It is usually interpreted as evidence either for
existence of some exotic matter components or for modification of the gravitational theory. In
the first route of interpretation one can take a perfect fluid with a sufficiently negative pressure,
dubbed dark energy \cite{dark}, to produce the observed acceleration. In the second route, however,
one attributes the accelerating expansion to a modification of general relativity.\\
This puzzling picture has been enhanced by the so-called Hubble tension \cite{hubble}. It consists of a discrepancy between the local value of the Hubble constant $H_0$ based on Cepheids in the Large Magellanic Clouds $H_0=74.03\pm 1.42~ Km s^{-1}Mpc^{-1}$ \cite{reiss} and the Planck data of the CMB radiation $H_0=67.4\pm 0.5~ Km s^{-1}Mpc^{-1}$ \cite{agh}.  Improved determination of the Hubble constant from the Planck observations and the Hubble Space Telescope measurements seem to demonstrate that this tension may not be caused by systematics \cite{sys}.
Hence, many authors prefer to believe that the tension might be caused by new physics beyond the $\Lambda$CDM.\\
These observations suggest that $H_0$ as inferred by CMB coming from $z\simeq 1100$ is lower than that inferred by the nearby sources. It therefore seems that the Hubble constant measurement could be affected by redshift in its
determination. A recent interesting analysis presented in \cite{pan} uses the Pantheon sample \cite{pan2}, the largest compilation of SNe Ia, to show that there is an evolution in the Hubble constant which scales as $g(z)=H_0 (1+z)^{-\delta}$ with $\delta$ and $H_0$ being the evolution parameter and the Hubble parameter at $z=0$, respectively. They divided the Pantheon sample into some bins where each bin is populated with an equal number of SNe Ia. Then they have fitted the values obtained in these bins with $g(z)$ fitting function with $\alpha\sim 10^{-2}$. This suggests that $H_0$ can be considered as a monotonic decreasing function from $z=0$ up to $z\simeq 1100$.\\
The possibility of a slow decaying of $H_0$ with redshift opens theoretical scenarios for its physical and dynamical interpretation. A possible proposition is
the modified gravity theories such as Brans-Dicke (BD) theory \cite{bd}, as a prototype of the scalar-tensor theories \cite{st},  or $f(R)$ gravity models \cite{modi}. It is well-known that $f(R)$ gravity, at least within a classical
perspective, are dynamically equivalent to the class of BD theories with a potential function and a null BD parameter \cite{sot}. Thus under a suitable redefinition of the gravitational and matter fields, one can make their field equations coincide. We will use this dynamical equivalence to consider Hubble tension in  a generalized BD theory (GBD) and $f(R)$ gravity in the Einstein frame and under the same analysis framework. We will use an exponential potential in this analysis. The choice of this kind of potential in $f(R)$ gravity is equivalent to considering a power-law $f(R)$ gravity \cite{pow}. We then determine the Hubble flow and show that even though decaying of $H(z)$ with redshift is possible for both models GBD is more consistent with the $g(z)$ model proposed in \cite{pan}.

~~~~~~~~~~~~~~~~~~~~~~~~~~~~~~~~~~~~~~~~~~~~~~~~~~~~~~~~~~~~~~~~~~~~~~~~~~~~~~~~~~~~~~~~~~~~~~~~~~~~~~~~~~~~~~~~~~~~~~~~~~~~~~~~
\section{The Model}
Let us begin with the following action\footnote{We use units in which $8\pi G=M_p^{-2}=1$.}
\begin{equation}
S^{BD}_{JF}=\frac{1}{2}\int d^4x \sqrt{-\bar{g}} \{\phi
\bar{R}-\frac{\omega_{BD}}{\phi}\bar{g}^{\mu\nu}\bar{\nabla}_{\mu}\phi
\bar{\nabla}_{\nu}\phi -V(\phi)+2 L_m\}
\label{1}\end{equation} where $\bar{R}$ is the Ricci scalar, $\phi$
is a scalar field, $V(\phi)$ is a potential function and $L_m$ is the matter Lagrangian density. The above action is the Jordan frame representation of the BD theory with the BD parameter $\omega_{BD}$ and a self-interacting potential $V(\phi)$. The potential generalizes the BD theory which is shown to be relevant for studying some cosmological problems \cite{v}.  \\
A conformal transformation
\begin{equation}
\bar{g}_{\mu\nu}\rightarrow g_{\mu\nu}=\phi ~\bar{g}_{\mu\nu}
\label{a2}\end{equation} brings the
above action into the Einstein frame \cite{far1}.  Then by a redefinition of the scalar field
\begin{equation}
\varphi=\sqrt{\omega_{BD}+\frac{3}{2}}~\ln
\phi \label{a3}\end{equation} the kinetic term takes a canonical form.  In terms of the new
variables ($g_{\mu\nu}$, $\varphi$) the action (\ref{1}) takes then
the form
\begin{equation}
S^{BD}_{EF}= \int d^{4}x \sqrt{-g} \{\frac{1}{2}R
-\frac{1}{2}g^{\mu\nu}\nabla_{\mu}\varphi
\nabla_{\nu}\varphi-U(\varphi)+e^{-\sigma\varphi}L_{m}\}
\label{a5}\end{equation} where
\begin{equation}
\sigma=8\sqrt{\frac{\pi}{2\omega_{BD}+3}}
\label{aaa5}\end{equation}
 and $\nabla_{\mu}$ is the covariant derivative of the
rescaled metric $g_{\mu\nu}$.  The Einstein frame potential is
$
U(\varphi)= V(\phi(\varphi))~e^{-\sigma\varphi}$.
 Now variation of the action (\ref{a5}) with
respect to the metric $g_{\mu\nu}$ and $\varphi$ gives,
respectively,
\begin{equation}
G_{\mu\nu}= (e^{-\sigma\varphi}T^{m}_{\mu\nu}+T^{\varphi}_{\mu\nu})
\label{2}\end{equation}
\begin{equation}
\Box \varphi-\frac{dU(\varphi)}{d\varphi}=\sigma e^{-\sigma\varphi} L_m \label{3}\end{equation}
where
\begin{equation}
T^{\varphi}_{\mu\nu}=(\nabla_{\mu}\varphi
\nabla_{\nu}\varphi-\frac{1}{2} g_{\mu\nu}\nabla_{\alpha}\varphi
\nabla^{\alpha}\varphi) -U(\varphi)g_{\mu\nu}
\label{4}\end{equation}
\begin{equation}
T^m_{\mu\nu}=\frac{-2}{\sqrt{-g}}\frac{\delta (\sqrt{-g}L_m)}{\delta
g^{\mu\nu}} \label{5}\end{equation}
Applying the Bianchi identities to (\ref{2}) gives
\begin{equation}
\nabla^{\mu}(e^{-\sigma\varphi}T_{\mu\nu}^m)=-\nabla^{\mu}T_{\mu\nu}^{\varphi}
 \label{3a}\end{equation}
which means that $T_{\mu\nu}^m$ and $T_{\mu\nu}^{\varphi}$ do not separately conserved. Here we
consider a perfect fluid energy-momentum tensor as a matter system
\begin{equation}
T^m_{\mu\nu}=(\rho_m+p_m)u_{\mu}u_{\nu}+p_mg_{\mu\nu}
\label{b1}\end{equation}
where $\rho_m$ and $p_m$ are energy density and pressure, respectively. The four-velocity of
 the fluid is denoted by $u_{\mu}$.
Details of
the energy exchange between matter and $\varphi$ depends on the
explicit form of $L_m$.
There are different choices for $L_m$
which all of them leads to the same energy-momentum tensor and field
equations in the context of general relativity \cite{haw} \cite{sh}.
Here we take $L_m=p_m$ for the
lagrangian density.\\
The $f(R)$ gravity models propose a modification of the Einstein-Hilbert action so that the Ricci scalar is replaced by some arbitrary function $f(R)$. There is a relation between the BD theory and $f(R)$ gravity in metric formalism at the classical level. In fact
metric $f(R)$ gravity is dynamically equivalent to GBD with a null BD parameter $\omega_{BD}=0$ \cite{sot}. It means that they are described by the same action and the resulting field equations coincide. The action for a metric $f(R)$ gravity in the Jordan frame is written as
\begin{equation}
S^{f(R)}_{JF}=\frac{1}{2}\int d^4x \sqrt{-g} \{f(R)+2L_m\}
\end{equation}
It can be easily shown that this action has a scalar-tensor representation \cite{sot} \cite{far}
\begin{equation}
S^{f(R)}_{JF}=\frac{1}{2}\int d^4x \sqrt{-g} \{\chi R-V(\chi)+2L_m\}
\label{aaa}\end{equation}
where the scalar field is defined as $\chi=f'(R)$. This is the BD theory in Jordan frame with $\omega_{BD}=0$. A conformal transformation $g_{\mu\nu}\rightarrow \Omega^2 g_{\mu\nu}$ with $\Omega=\sqrt{\chi}=f'^{1/2}(R)$ and the scalar field redefinition $\xi=\sqrt{\frac{3}{2}}\ln\chi$\footnote{This is the field redefinition (\ref{a3}) with $\omega_{BD}=0$.} bring the action (\ref{aaa}) into the Einstein frame representation
\begin{equation}
S^{f(R)}_{EF}= \int d^{4}x \sqrt{-g} \{\frac{1}{2}R
-\frac{1}{2}g^{\mu\nu}\nabla_{\mu}\xi
\nabla_{\nu}\xi-U(\xi)+
e^{-\sqrt{\frac{2}{3}}\xi}L_{m}\}
\label{aa5}\end{equation}
where
\begin{equation}
U(\xi)=\frac{Rf'(R)-f(R)}{2f'^2(R)}
\end{equation}
with prime being derivative with respect to $R$. Variations of (\ref{aa5}) with respect to $g^{\mu\nu}$ and $\xi$ gives the same field equations (\ref{2}), (\ref{3}) and (\ref{4}) when $\varphi$ is replaced by $\xi$ and $\sigma$ takes the value $\sigma= 8\sqrt{\frac{\pi}{3}}\simeq 8.2$.\\
As the last point in this section we comment that although Solar System experiments set the constraint $\omega_{BD}>40000$ \cite{will}, the null BD parameter does not generally mean that $f(R)$ gravity models do not satisfy observational constraints. In fact, some authors have already argued that since the post-Newtonian parameter satisfies $\gamma=\frac{1}{2}$ instead of being equal to unity as required by observations, all f(R) theories should be ruled out \cite{chiba}.  Later, it was noted that
for scalar fields which have sufficiently large masses it is possible for $\gamma$ to be close to unity even for null
BD parameter. In this case the scalar fields become short-ranged and have no effect
at Solar System scales \cite{pow}.

~~~~~~~~~~~~~~~~~~~~~~~~~~~~~~~~~~~~~~~~~~~~~~~~~~~~~~~~~~~~~~~~~~~~~~~~~~~~~~~~~~~~~~~~~~~~~~~~~~~~~~~~~~~~~~~~~
\section{Cosmological implementation}
 We apply the field equations (\ref{2}) and
(\ref{3}) to a spatially flat Friedmann-Robertson-Walker spacetime
$
ds^2=-dt^2+a^2(t)(dx^2+dy^2+dz^2)$ with
$a(t)$ being the scale factor.  This gives
\begin{equation}
3H^2=e^{-\sigma\varphi}\rho_{m}+\rho_{\varphi}
\label{a11}\end{equation}
\begin{equation}
2\dot{H}+3H^2=-(e^{-\sigma\varphi}p_{m}+p_{\varphi})
\label{a12}\end{equation}
\begin{equation}
\ddot{\varphi}+3H\dot{\varphi}+\frac{dU(\varphi)}{d\varphi}=-\sigma e^{-\sigma\varphi}p_{m}
\label{a13}\end{equation}
where $H=\frac{\dot{a}}{a}$ is the Hubble parameter. The conservation equations
become
\begin{equation}
\dot{\rho}_{\varphi}+3H(\omega_{\varphi}+1)\rho_{\varphi}=-\sigma e^{-\sigma\varphi}\dot{\varphi}\rho_m
\label{a15}\end{equation}
\begin{equation}
\dot{\rho}_{m}+3H(\omega_m+1)\rho_m=(1-2\omega_m)\sigma  \dot{\varphi}
\rho_m \label{a14}\end{equation}
 The latter can be solved which gives the following solution
\begin{equation}
\rho_m=\rho_{0m} a^{-3(\omega_m+1)}e^{(1-2\omega_m)\sigma\varphi}\label{1a16}\end{equation} where
$\rho_{0m}$ is an integration constant. This solution can also be written as
 \cite{bis1}
\begin{equation}
\rho_m=\rho_{0m} a^{-3(\omega_m+1)+\varepsilon}
\label{a166}\end{equation} where we have defined
\begin{equation}
\varepsilon\equiv \lambda(1-2\omega_m)\frac{\sigma\varphi}{\ln
a}\label{a16}\end{equation}
with $\lambda$ being a parameter. This solution indicates that the
evolution of energy density is modified due to interaction of
$\varphi$ with matter.  For $\varepsilon>0$, matter
is created. In this case, energy is injecting from $\varphi$ into the matter so that the latter dilutes more slowly compared to the standard evolution
$\rho_m\propto a^{-3(\omega_m+1)}$. For $\varepsilon<0$, on the other hand, matter is annihilated and energy transfers outside of the matter system. In this case, the rate of dilution of $\rho_m$ is faster than the standard one.\\
We now search for solutions for the field equations (\ref{a11}), (\ref{a12}) and (\ref{a13}) which are able to reproduce the current
Universe expansion and can specifically alleviate the tensions in the Hubble constant measurements. In particular, we are searching for those solutions for which $H(z)$ receives variations with the redshift as a consequence of evolution of the non-minimally coupled scalar field $\varphi(z)$. There are two independent equations among (\ref{a11})-(\ref{a13}) which ultimately give $H(z)$ and $\varphi(z)$. Moreover, the conservation equations (\ref{a15}) and (\ref{a14}) characterize the energy exchange between $\varphi$ and matter which modify evolution of $\rho_m$ and $\rho_{\varphi}$. In this interacting system, $\varepsilon$ in (\ref{a166}) is generally an evolving function characterizing the rate of the energy transfer. However, we consider
the case that $\varepsilon$ can
be regarded as a constant parameter. Even though this is not generally true during the whole expansion history of the Universe, it may hold during particular eras or during short periods in the evolution of the Universe. In this case, (\ref{a16}) reduces to
\begin{equation}
\varphi=\frac{\varepsilon}{\lambda\sigma(1-2\omega_m)} \ln a
\label{c2}\end{equation}
which implies that the rate of change of the scalar field is given by the Hubble parameter, namely $\dot{\varphi}\propto H$. Now we study the Friedman equation (\ref{a11}) for GBD theory and $f(R)$ gravity: \\
\emph{1) GBD theory.} As the simplest and the best-studied generalization of general relativity, it is natural to think about
the BD scalar field as a possible candidate for producing modifications of cosmic expansion. We first consider the potential $V(\phi)=V_0 e^{\alpha\phi}$ in the action (\ref{1}) with $V_0$ and $\alpha$ being constants. This is equivalent to $U(\varphi)=V_0 e^{(\alpha-\sigma)\varphi}$ in the action (\ref{a5}). In this case, the Friedman equation (\ref{a11}) becomes
\begin{equation}
\frac{H(z)}{H_0}=(3-\frac{\varepsilon^2}{2\lambda^2 \sigma^2})^{-1}\{3\Omega_{0m} (1+z)^{-3+\varepsilon(1-\frac{1}{\lambda})}+(1+z)^{\frac{\varepsilon}{\lambda\sigma}(\alpha-\sigma)}\}^{\frac{1}{2}}
\label{b33}\end{equation}
\begin{figure}[ht]
\begin{center}
\includegraphics[width=0.4\linewidth]{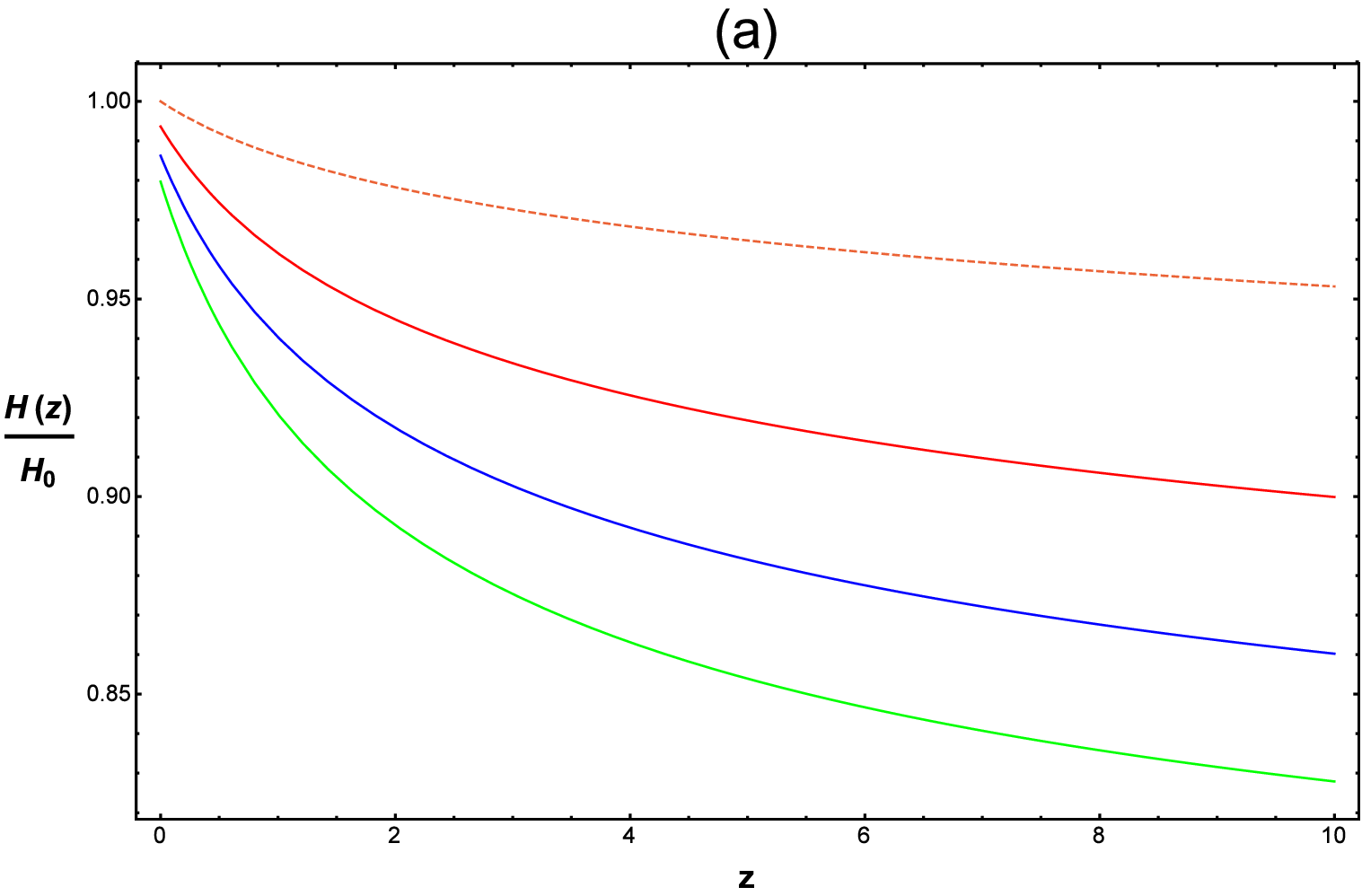}
\includegraphics[width=0.4\linewidth]{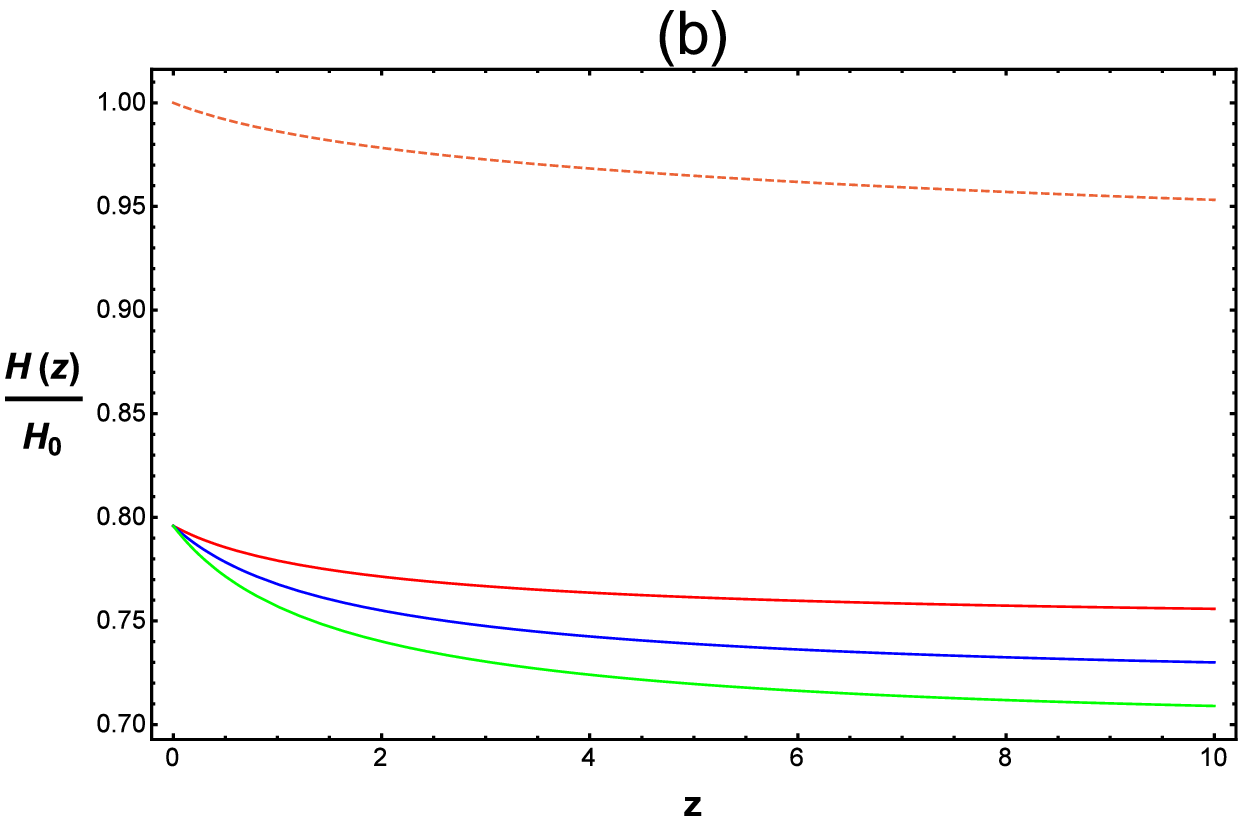}
\caption{The plots of $H(z)/H_0$ in BD theory with an exponential potential (panel a) and $f(R)$ gravity (panel b).  In these plots $\omega_m=0.3$; $\alpha=0.04$; $\varepsilon=3.3$, 3.4, 3.5 and $\lambda=45, 47, 49$ from top to bottom. The dashed line denotes the fitting function $g(z)/H_0$ used in \cite{pan}. }
\end{center}
\end{figure}
where we have set $V_0=H_0^{2}$ and $\Omega_m\equiv\rho_{0m}/\rho_c$, $\rho_c\equiv 3H^2_0$ and $a=(1+z)^{-1}$ are used\footnote{We have used $\frac{a_0}{a}=1+z$ with $a_0=1$.}. Here $H_0$ is taken as the late-time Hubble constant or that one measured by the SNe Ia. In this equation, $\sigma<0.05$ which is inferred from the observational bound $\omega_{BD}>40000$ and (\ref{aaa5}). The function $H(z)/H_0$ is plotted in Fig. 1(a). The figure shows that the non-minimally coupled scalar field is actually
responsible for scaling of $H(z)$. Interestingly, this scaling behaviour is similar to that of the fitting function $g(z)/H_0$ (the dashed line) used in the analysis of the Pantheon sample in \cite{pan}. Both functions increase slowly with decreasing of $z$ and go to $H_0$ at late times. This means that the GBD theory is a good theoretical interpretation for the model $g(z)=H_0 (1+z)^{-\delta}$ with $\delta=0.02$. \\
\emph{2) f(R) Gravity.} The Hubble parameter $H(z)$ in the $f(R)$ gravity models is also given by (\ref{b33}) when $\sigma$ takes the value $\sigma\simeq 8.2$. In this case, exponential potentials correspond to power-law $f(R)$ gravity \cite{pow}. The function $H(z)/H_0$ is plotted in Fig. 2(b).
The figure shows that although $H(z)$ in power-law $f(R)$ gravity has the same scaling behaviour as GBD theory, it does not goes to $H_0$ at late times.  Thus GBD theory provides a better theoretical framework for understanding discrepancies in the Hubble constant measurements.
~~~~~~~~~~~~~~~~~~~~~~~~~~~~~~~~~~~~~~~~~~~~~~~~~~~~~~~~~~~~~~~~~~~~~~~~~~~~~~~~~~~~~~~~~~~~~~~~~~~~~~~~~~~~~~
\section{Conclusions}
In this work, we have considered the possibility that the Hubble constant measurements are affected by redshift so that measurements based on CMB at $z\simeq 1100$ are smaller than those based on local objects such as Cepheids in the Large Magellanic Clouds and SNe Ia. This possibility is supported by analyses  such as \cite{pan} \cite{gal} which fitted extracted $H_0$ values with a function mimicking the redshift evolution $g(z)\propto (1+z)^{-0.02}$. A theoretical route of interpretation of this observation is the modified theories of gravity.\\
We have used the dynamical equivalence of $f(R)$ gravity and GBD with null BD parameter to investigate Hubble tension under the same analysis framework. In this analysis a pure exponential potential is used which corresponds to a power-law $f(R)$ gravity. The two class of theories are described by the same set of field equations in Einstein frame so that they are distinguished by the parameter $\sigma$. For GBD theories, $\omega_{BD}>40000$ corresponds to the upper bound $\sigma<0.05$ and for $f(R)$ gravity the null BD parameter is equivalent to $\sigma\simeq 8.2$. We have shown that there is a monotonic decreasing trend in the Hubble constant $H_0$ in both theories which mostly happens for $z< 2$. However, there are two weakness of the $f(R)$ gravity: First, in order that power-law $f(R)$ gravity be consistent with local gravity experiments the exponent of the curvature scalar hardly deviates from unity. Second, contrary to GBD the asymptotic behaviour of $H_0$ at late times is not consistent with that measured from the local distance ladder.

\end{document}